\documentstyle[11pt,psfig]{article}
\newcommand{\eqn}[1]{Eq.~(\ref{eqn:#1})}
\newcommand{\fig}[1]{Fig.~\ref{fig:#1}}
\newcommand{\se}[1]{section~\ref{sec:#1}}

\newcommand{\la}{\langle}
\newcommand{\ra}{\rangle}
\newcommand{\La}{\left\langle}
\newcommand{\Ra}{\right\rangle}
\newcommand{\gsim}{\stackrel{>}{\sim}}

\newcommand{\Rg}{R_{\rm g}}
\newcommand{\tm}{\tau_{\rm m}} 
\newcommand{\tp}{\tau_{\rm p}}
\newcommand{\tr}{\tau_{\rm rep}}
\newcommand{\tR}{\tau_{\rm R}}
\newcommand{\dc}{d_{\rm c}}
\newcommand{\Dk}{D_1} 

\newcommand{\Dz}{D_{\rm z}}
\newcommand{\Dxy}{D_{\rm xy}}
\newcommand{\DR}{D_{\rm R}}
\newcommand{\DE}{D_{\rm E}}

\newcommand{\CR}{\cal{R}}
\newcommand{\CRE}{\cal{R}_{\rm E}}
\newcommand{\CRR}{\cal{R}_{\rm R}}
\newcommand{\CC}{\cal{C}}
\newcommand{\CCE}{\cal{C}_{\rm E}}
\newcommand{\CCR}{\cal{C}_{\rm R}}
\newcommand{\Ne}{N_{\rm e}}
\newcommand{\bigO}{\cal{O}}

\title{Dynamics of a Polymer in the Presence of Permeable Membranes}
\author{Hyoungsoo Yoon \quad J.M. Deutsch \\
        University of California, Santa Cruz \\
        Santa Cruz, CA 95064}

\begin{document}
\maketitle

\begin{abstract}

We study the diffusion of a linear polymer in the presence of
permeable membranes without excluded volume interactions, using
scaling theory and Monte Carlo simulations.  We find that the average
time it takes for a chain with polymerization
index~$N$ to cross a single isolated membrane varies with~$N$ as~%
$N^{5/2}$, giving its permeability proportional to~$N^2$.  When the
membranes are stacked with uniform spacing~$d$ in the unit of the
monomer size, the dynamics of a polymer is shown to have three
different regimes.  In the limit of small~\mbox{$d \ll N^{1/2}$}, the
chain diffuses through reptation and \mbox{$D\sim N^{-2}$}.  When $d$
is comparable to~$N^{1/2}$ the diffusion coefficients parallel and
perpendicular to the membranes become different from each other.
While the diffusion becomes Rouse-like, i.e.~\mbox{$D\sim N^{-1}$}, in
the parallel direction, the motion in the perpendicular direction is
still hindered by the two-dimensional networks.  The diffusion
eventually becomes isotropic and Rouse-like for large~\mbox{$d\gg N$}.

\end{abstract}

\section{Introduction}
\label{sec:introduction}

The physics of tethered membranes%
\cite{YKMKetal:StatisticalMechanics,%
EGFDetal:ThermodynamicalBehavior,%
IBPGSG:MolecularDynamics,%
RL:StatisticalPhysics}
has been studied with great
interest due not only to its relevance to biology but also to its
numerous applications in technology.
Stabilizing liposomes, which are used in drug delivery, by tethered
membranes is just one of many important applications%
\cite{HRBSetal:MolecularArchitecture}.
It is also possible that tethered membranes could provide 
a new way of
filtrating polymers {\em in a solution\/}
according to their molecular weights.
This method should be especially useful where conventional methods
such as gel electrophoresis%
\cite{MOdlCJMDetal:ElectrophoresisPolymer,%
JMDTLM:TheoreticalStudies,%
SPOMR:ScalingMegabase}
do not work.

Motivated by this, we study
in this paper the effects of permeable membranes on the
dynamics of a flexible linear polymer.
More specifically we focus on the
diffusion characteristics of a linear polymer 
in the presence of a single or multiple membranes.

We model our permeable membranes as two-dimensional
regular networks of polymers with
the mesh size bigger than the persistence length of the
diffusing chain.
This type of membranes can be made by crosslinking linear polymers
using ultraviolet light\cite{HRBSetal:MolecularArchitecture}
or by extracting spectrin networks from human
red blood cells\cite{CFSKSetal:ExistenceFlat}.
The widely used molecular filters, called zeolites%
\cite{RFS:MakingMolecular}, 
may also provide a good system
if the molecular parameters such as the Kuhn length of the
diffusing polymer is sufficiently small.

For simplicity, we restrict our discussion to systems with no excluded
volume interactions.
The only constraints the two-dimensional networks impose on the
dynamics of the chain polymer are topological barriers.
From a theoretical point of view, this is a new problem
quite distinct from the 
study of reptating chains in an inhomogeneous environment%
\cite{ITKHLetal:ReptationDynamics,BHZOL:ReptationPolymer},
which has
been investigated extensively partly due to its relevance to DNA gel
electrophoresis.
Even though the main ingredients in reptation theory 
are the topological
constraints imposed by surrounding chains or gel networks, 
most studies on reptation in
an inhomogeneous medium concentrates on more
complex (and presumably {\em more realistic\/}) 
systems with excluded volume interactions.

The remainder of this paper is organized as follows.
In \se{model} we first define our problem in more precise terms and
give a brief account of the simulation methods used in this work.
In \se{relaxation_time}, 
we consider 
the longest relaxation time of a linear
polymer in the presence of a {\em single\/} membrane.
This is obtained numerically by measuring
the first passage times of the linear
polymer through an array of membranes which are placed with
a spacing proportional to the radius of gyration of the polymer.
In \se{permeability}, we use the data obtained in the previous section
to calculate the permeability for our
two-dimensional network, which is one of the most important
characteristics of permeable membranes.
In \se{diffusion_constants}, we study the multiple membrane case
with uniform spacings in more detail.
The diffusion constants parallel to the membranes
as well as the perpendicular ones are studied.
We first present a simple scaling theory as a function of the
polymerization index and the spacing between the membranes.
Then we present the simulation results 
which support this analysis.
In \se{escape}, we consider
the mean square displacements of a monomer as a function of time
when a chain disengages from a network.
Finally in \se{conclusion}
we summarize our results and discuss possible applications of
this study.

\section{The Model}
\label{sec:model}

We consider a random walk chain of $N$ monomers or beads on a
cubic lattice with a lattice spacing~$b$, which is
the monomer size or the bond length of the chain.
We only consider
even~$N$'s due to the bipartite nature of the cubic lattice.

The simulation is done by allowing the chain to move
according to the following set of Monte Carlo rules
which mimics Rouse dynamics.
\begin{itemize}
\item[I] Two consecutive bonds which form an $L$-shape can be inverted,
  in the plane formed by the~$L$.
\item[II] A {\em kink}, two consecutive bonds which are on top of each other,
  can be flipped to one of the six lattice directions including the original
  one. (See, for example, Fig.~1 of 
  reference~\cite{JMDHY:SimulationForces}.)
  The free ends of the chain are treated as kinks.
\end{itemize}

Now we consider another cubic lattice with lattice spacing~$a$
which interpenetrates the first one.
We assume $a$ is an integral multiple of~$b$ so that
the lattice points of the second lattice always reside at the centers
of the unit cells of the first.
We let $a=b=1$ in the numerical simulations.

Our membrane is a two-dimensional
slice of the second lattice in the horizontal plane.  Since the
membrane occupies a different lattice than that of the chain, it does
not alter the statics of the chain. But we assume that the chain and
the membranes cannot pass through each other.
That is, the links of the
membranes act as topological barriers for the motion of the chain.

This is implemented in our simulation by modifying rule~I. We
allow the $L$-inversion only when the chain does not pass through a link
of a membrane by this move.
This is basically the same as the Doi-Evans-Edwards algorithm%
\cite{MD:CageModel,KEESFE:ComputerSimulation:1}.
Note that 
the dynamics of a chain is altered only near the vicinity of
the membranes and
if the plane formed by the~$L$ is horizontal rule~I is
not affected at all
because the links of the membranes are all horizontally placed.

\section{Relaxation time of chain penetration}
\label{sec:relaxation_time}

First we study the effects of a single membrane on the
dynamics of a linear polymer with polymerization index~$N$.
One of the most basic questions we have to ask in this situation is
how much the membrane will slow down the diffusion of the polymer
which is in its vicinity.
More specifically we ask what is the time scale of the complete
penetration of the polymer through the 
two-dimensional network.

Before we define the relaxation time for this penetration, we point out
some problems associated with the very concept of the
``complete penetration''.
First we note that 
because a polymer has a finite size of order the radius of gyration,
\mbox{$\Rg \sim N^{1/2}b$},
the presence of the membrane interferes with the Rouse dynamics of the
chain as long as its center of mass lies in a slab-like region
around the membrane.
We call this region of thickness~${}\sim \Rg$ 
the {\em entanglement zone}.
Here we stress that the thickness of the entanglement
zone is a function of~$N$.

If we try to define the relaxation time of a chain
through a membrane by
diffusion, we soon encounter difficulties.
Diffusion is a random process and 
the polymer may never pass through the
membrane. 
If the polymer penetrates the membrane at some point in time, the
question remains as to how much time should be attributed to the
effect of the membrane.

We circumvent this problem in the following way.
First we consider two configurations, one with the center of mass of
the polymer at a distance~${}\sim \Rg$ from the membrane 
and another in which the polymer is {\em maximally entangled\/} 
(on average) with
the membrane, that is, its center of mass lies at the membrane.
Since the equilibrium statics are not affected by the topological
membrane, these two configurations should arise with the same
probability in equilibrium.
Then from the principle of detailed balance, the
transition probability from one configuration to the other should be
the same as the reverse.
Hence the transition times in both directions are the same.

If we place a chain in the maximally entangled position 
(as defined in the previous paragraph)
the chain
eventually diffuses out to {\em either\/} end of the entanglement zone.
We define the average first passage time in this process as
the {\em  escape time}.
Then we define the {\em penetration time\/} as (4 times) the escape time
since, as described above, the escaping process takes the same amount
time as the reverse process and the penetration is, by definition, the
combination of the two.
Note that the chain does {\em not necessarily\/} penetrate to the
other side of the membrane in this time scale.

The penetration time defined this way can be estimated easily by
simple application of the reptation theory%
\cite{PGdG79:ScalingConcepts}.
First we note that the chain typically crosses the membrane 
\mbox{${}\sim N^{1/2}$}~times 
in the entanglement zone with the average contour
length~\mbox{${}\sim N^{1/2}b$} between the crossing points, or
the {\em entanglement points}.
The relaxation of the chain can be divided into two processes; Rouse
relaxation of chain segments between entanglement points, which we
call {\em Rouse segments}, and diffusion of kinks past entanglement
points.

Even though this situation is quite different from the one
of the usual
reptation theory
described by ``reptilian'' motion in a
tube, we can still imagine the chain wiggling through a series
of entanglement points inside an imaginary tube,
its mean diameter being of order~$N^{1/4}b$.

The longest relaxation time in reptation theory
is $\tau_1 N^3/\Ne$, 
where $\Ne$ is the number of monomers between entanglement points.
Since $\Ne \sim N^{1/2}$ in our case, 
we obtain\cite{PGdG79:ScalingConcepts} 
\begin{equation}
  \tm \sim N^{5/2} \tau_1
\end{equation}
Therefore
dynamics of a chain through a tethered membrane is 
faster than the reptation in a three-dimensional network
\mbox{$\tr \sim N^3 \tau_1$} 
and slower than the Rouse diffusion 
\mbox{$\tR \sim N^2 \tau_1$}, as expected.
Note that $N^{5/2}$-dependence has been known to exit in various
contexts such as the dynamics of a ring polymer in a gel%
\cite{SPOMRetal:DynamicsRing}.

In a time scale of $\tm$ the chain moves a distance of the
order~$\Rg\sim N^{1/2}b$. 
Hence if we place an infinite array of membranes with
${}\sim \Rg$ apart from one another then the average time for a
chain to hop across a membrane should be proportional to the
relaxation time~$\tm$.
We conducted a simulation under these conditions for chain lengths 8
through~80.
The inter-membrane distances are set to $2bN^{1/2}$, with $b=1$.
At these distances the chains touching two adjacent membranes
at the same time is highly unlikely.
The penetration time,~$\tau \sim \tm$ is numerically defined as follows.
\begin{equation}
\frac{1}{\tau} = \La \frac{1}{\tp} \Ra
\end{equation}
where $\tp$ is the first passage time from the center of one
inter-membrane zone to a neighboring one.
The angular brackets indicate the average 
over the distribution of~$\tp$'s.

\begin{figure}[p]
\begin{center}
\
\psfig{file=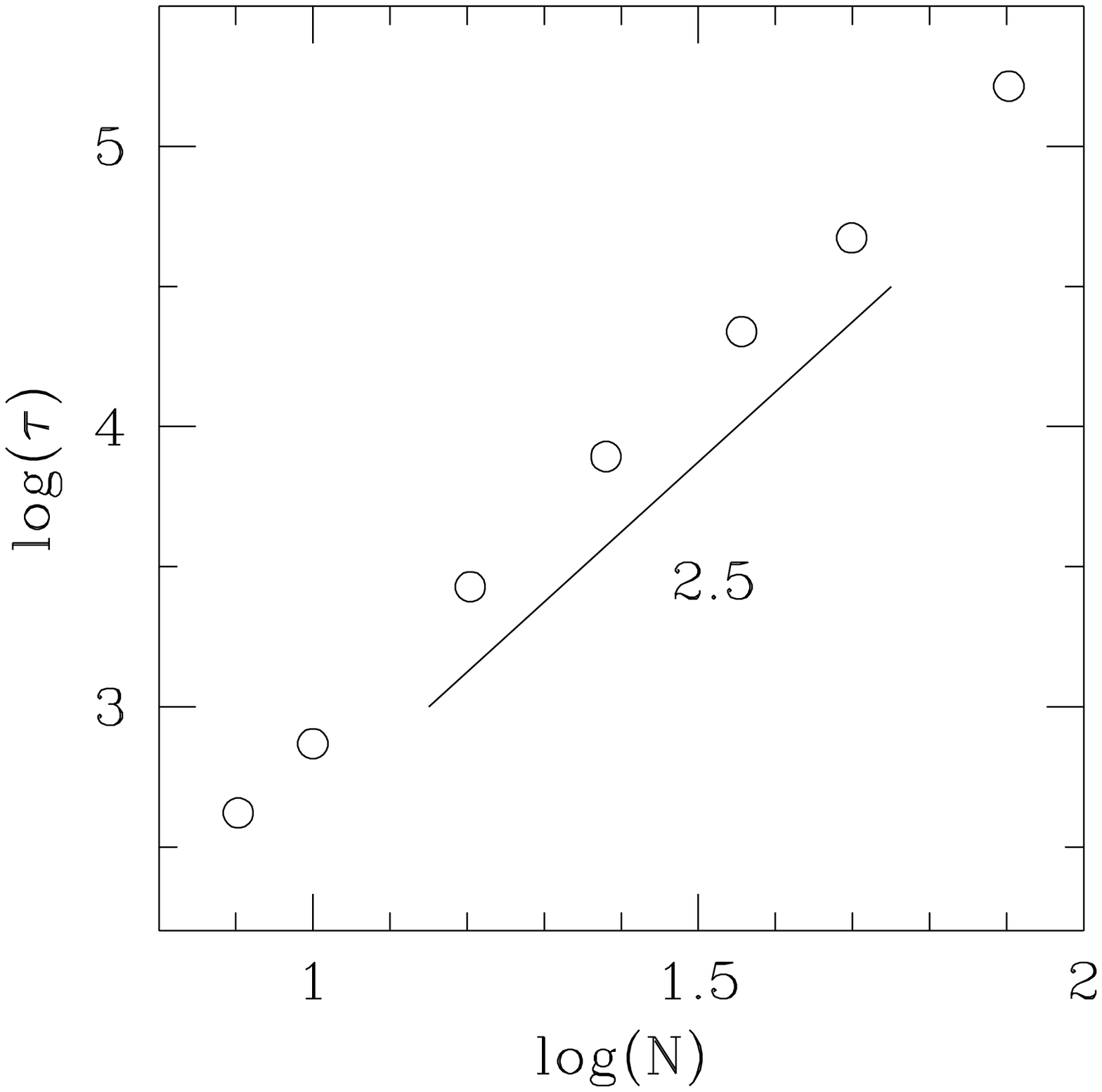,width=6in}
\end{center}
\caption[]{The logarithm of penetration time, $\tau$, defined in the
  text is plotted against the logarithm of the polymerization
  index,~$N$. The membranes are placed at distances 5, 6, 8, 10, 12,
  14 and~18 for $N=8$, 10, 16, 24, 36, 50 and~80. The distances are
  chosen so that the probability of the chain touching two membranes
  at the same time is extremely small.}
\label{fig:relaxation}
\end{figure}

The log-log plot of $\tau$ vs.~$N$ is
shown in~\fig{relaxation}.
The typical number of $\tp$'s used in obtaining
each point is approximately~$10^4$.
As the figure shows the exponent is clearly close to~$2.5$ as the above
scaling theory predicts.
If there were no fluctuations in the size of the chain, we should be
able to produce more accurate results by shrinking the intermembrane
distance down to the radius of gyration of the chain.
Because of the more generous spacing used in the simulation the
results must contain a part from Rouse dynamics.
But the effect should be small because $\tm \gg \tR$ for
large~$N$. 

The penetration time is also expected to
depend on the mesh size of the
membrane,~$a$. 
Even though
the $a$ dependence is harder to ascertain in our lattice models since $a$
only takes integer values,
it can be obtained using simple scaling argument as
follows. 
First we assume $\tm$ is a function of only $N$ and $a/b$. And since
we already know the $N$-dependence we write $\tm$ as
\begin{equation}
  \tm = \tau_1 N^{5/2}\left(\frac{a}{b}\right)^p
\end{equation}
where $p$ is the scaling exponent.
When $a$ approaches~$N^{1/2}b$ the dynamics should be
Rouse-like.
Hence we get $p=-1$.
That is, $\tm$ is linearly proportional to the inverse of the mesh
size~$a$ unlike~$\tr$, which has the quadratic dependence 
on~$a^{-1}$.

\section{Permeability of a membrane}
\label{sec:permeability}

As suggested earlier, two-dimensional networks of polymers or
two-dimensional slabs with pores could be used for
filtration purposes.
One of the most fundamental quantities in the membrane filtration
technique is the {\em permeability\/} of a membrane,~$P$, which is defined
by the following empirical equation,
\begin{equation}
J = - P \Delta c
\end{equation}
where $J$ is the flux density of the solute molecules 
through the membrane
and $\Delta c$ is
their concentration difference across the membrane.

A membrane is called {\em selectively permeable\/} 
when $P$ depends on the
solute being filtered.
Our model membrane is selectively
permeable to polymer solutes of different lengths
as will be shown below and this is the
basis for our proposal that tethered membranes be used as a dialysis
tool. 

From Fick's law we can derive $P$ in terms of quantities
which are more easily measurable, the diffusion constants.
\begin{equation}
  \frac{1}{P} = \int_{\rm entanglement\ zone}
  \frac{dz}{\Dz(z)}
  \label{eqn:permeability}
\end{equation}
where $\Dz$ is a diffusion constant in the perpendicular direction to
the membrane and
the thickness of the entanglement zone is of order~$\Rg$, 
as described earlier.
The diffusion constant in this entanglement zone should vary due to
the spherical shape of the polymer,
as indicated by the $z$-dependent $\Dz$ in \eqn{permeability}.
But we neglect this as in the previous section and define the average
diffusion constant along the $z$-direction,~$\bar{\Dz}$,
using the following formula
\begin{equation}
  \frac{\Rg}{\bar{\Dz}} \equiv \int\frac{dz}{\Dz(z)}
\end{equation}
Then from the definition of $\tm$ we get
\begin{equation}
  \bar{\Dz} = \Rg^2/\tm \sim N^{-\frac{3}{2}}
\end{equation}

Hence from \eqn{permeability} or
\begin{equation}
  \frac{1}{P} \approx \frac{\Rg}{\bar{\Dz}}
  \label{eqn:perm2}
\end{equation}
we get 
\begin{equation}
  P \sim N^{-2}
\end{equation}

\section{Diffusion constants of a polymer}
\label{sec:diffusion_constants}

Now we consider the multiple membrane case.
We stack membranes along the $z$-direction with a spacing~$d$.
A linear polymer of polymerization index~$N$ is introduced at an
arbitrary position in this system and the diffusion constants are
measured both along the $z$-direction and in the $xy$-plane 
using the following equations.
\begin{equation}
  \Dz = \lim_{t \rightarrow \infty} \frac{\la (Z(t)-Z(0))^2 \ra}{2t}
\end{equation}
and
\begin{equation}
  \Dxy = \lim_{t \rightarrow \infty} 
  \frac{\la (X(t)-X(0))^2 + (Y(t)-Y(0))^2 \ra}{4t}
\end{equation}
where $X, Y$ and $Z$ are center of mass coordinates of the chain.

First we expect that there should be a crossover
in the $D$'s at $\dc \sim \Rg$. 
When $d$ is smaller than $\dc$ the chain always entangles with multiple
membranes
and we expect that some sort of modified reptation
should occur.
On the other hand
when $d$ is much larger than~$\dc$ the membranes act as
small perturbations to the chain dynamics which should be
otherwise Rouse-like.
We study both limits in turn.

\subsection{$d \ll \dc$}

In this dense regime of membranes
we expect the reptation theory to be useful in
describing the dynamics of a linear polymer,
since the array of
closely placed membranes should not be much different from
a three-dimensional network.

We first estimate the density of entanglement points.
The polymer intersects of order $\Rg/d \sim N^{1/2}b/d$ membranes on
average. 
Furthermore
it crosses each membrane of order $\Rg/a \sim N^{1/2}b/a$ times,
which can be seen by rescaling the monomer size~$b$ to the mesh
size~$a$, so that $N \rightarrow Nb^2/a^2$.
Hence the total number of entanglement points is 
$\Rg^2/ad \sim Nb^2/ad$.
Then the average number of monomers between entanglement points
is obtained by dividing~$N$ by this number, that is, we obtain
$\Ne \sim ad/b^2$. 
Naive application of reptation theory%
\cite{PGdG79:ScalingConcepts} 
as in \se{relaxation_time}, yields
\begin{eqnarray}
  \tau &=& \tau_1 \frac{N^3b^2}{ad}, \\
  D &=& \Dk \frac{ad}{N^2b^2}
\end{eqnarray}
The diffusion constant~$D$ increases as a linear function of the
spacing~$d$. 

\begin{figure}[p]
\begin{center}
\
\psfig{file=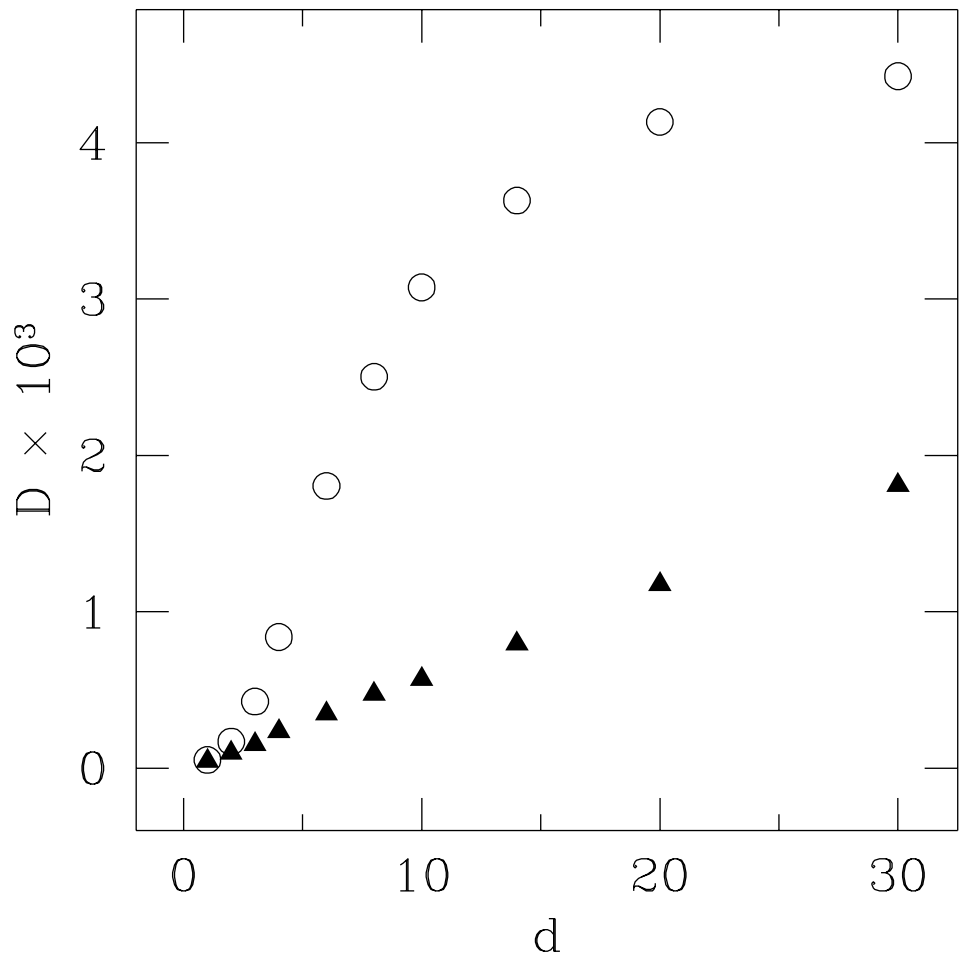,width=6in}
\end{center}
\caption[]{The diffusion constants are given as functions of~$d$ for
  $N=50$. The circles represent $\Dxy$ and the triangles~$\Dz$. The
  Rouse value is around $5.5\times 10^{-3}$ in our unit. Each pair of
  points represents an average over 1000 Monte Carlo runs.}
\label{fig:diffusion}
\end{figure}

The simulation results for $\Dz$ and~$\Dxy$
for $N=50$ are plotted in \fig{diffusion}
against~$d$ up to~$d=30$.
Each point in this figure was averaged over 1000 Monte Carlo runs,
each spanning longer than the penetration time.
As the above argument based on the reptation theory predicts,
diffusion should be isotropic in this regime.
As \fig{diffusion} shows, this is indeed the case in the limit of 
$d/\Rg \rightarrow 0$.
But the system becomes anisotropic rather rapidly as the spacing
increases. 
While $\Dz$ remains linear in~$d$ even well above~$\dc$,
$\Dxy$ deviates dramatically from a straight line
around a relatively small~$d$.

\begin{figure}[p]
\begin{center}
\
\psfig{file=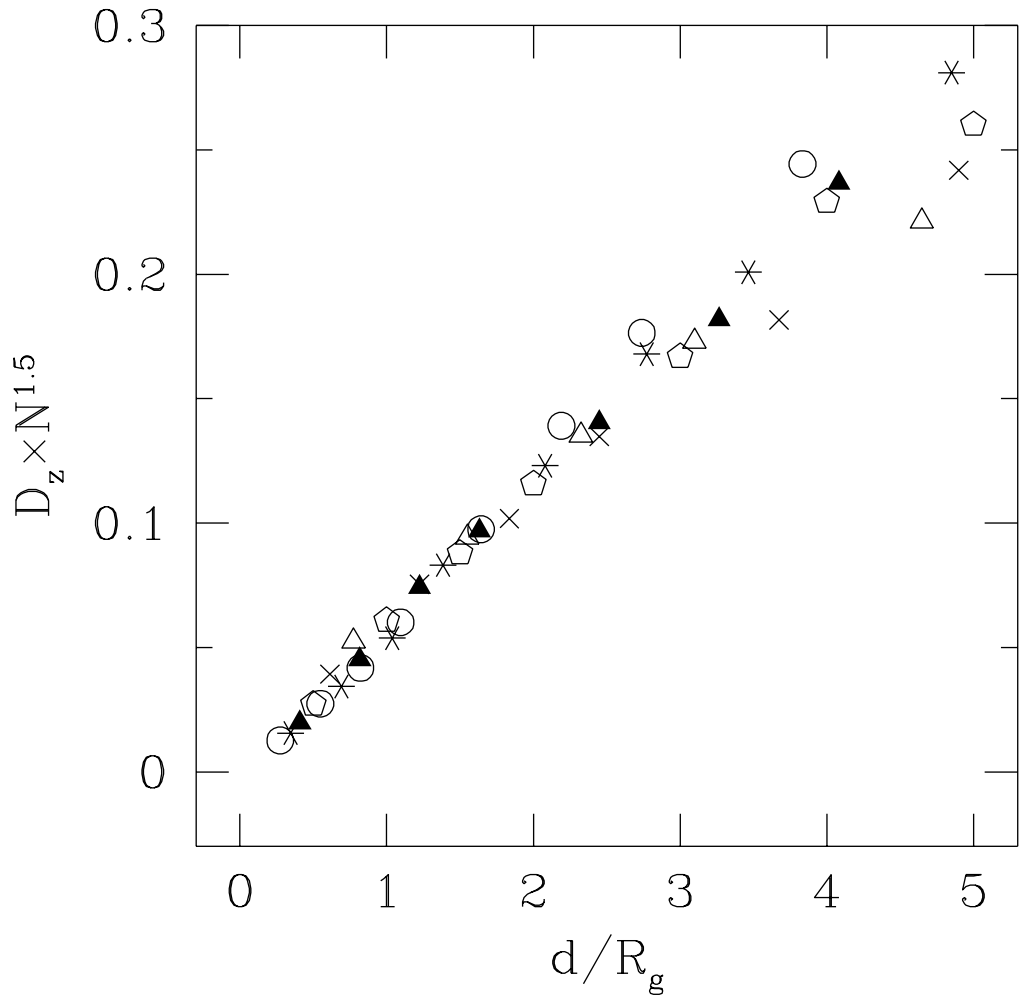,width=6in}
\end{center}
\caption[]{Scaling graph for $\Dz$ for $N=10$(open triangle),
  16(cross), 24(pentagon), 36(filled triangle), 50(snowflake) and
  80(circle). The scaling function is $\Dz = D_1/N^{1.5}F(d/\Rg)$
  where $\Rg = \sqrt{Nb^2/6}$ is used. $F$ is pretty linear when
  $d/\Rg$ is small and the scaling breaks down beyond that. Note that
  each data point was obtained separately. Each point is an average of
  1000 Monte Carlo runs.}
\label{fig:scaling_z}
\end{figure}

\begin{figure}[p]
\begin{center}
\
\psfig{file=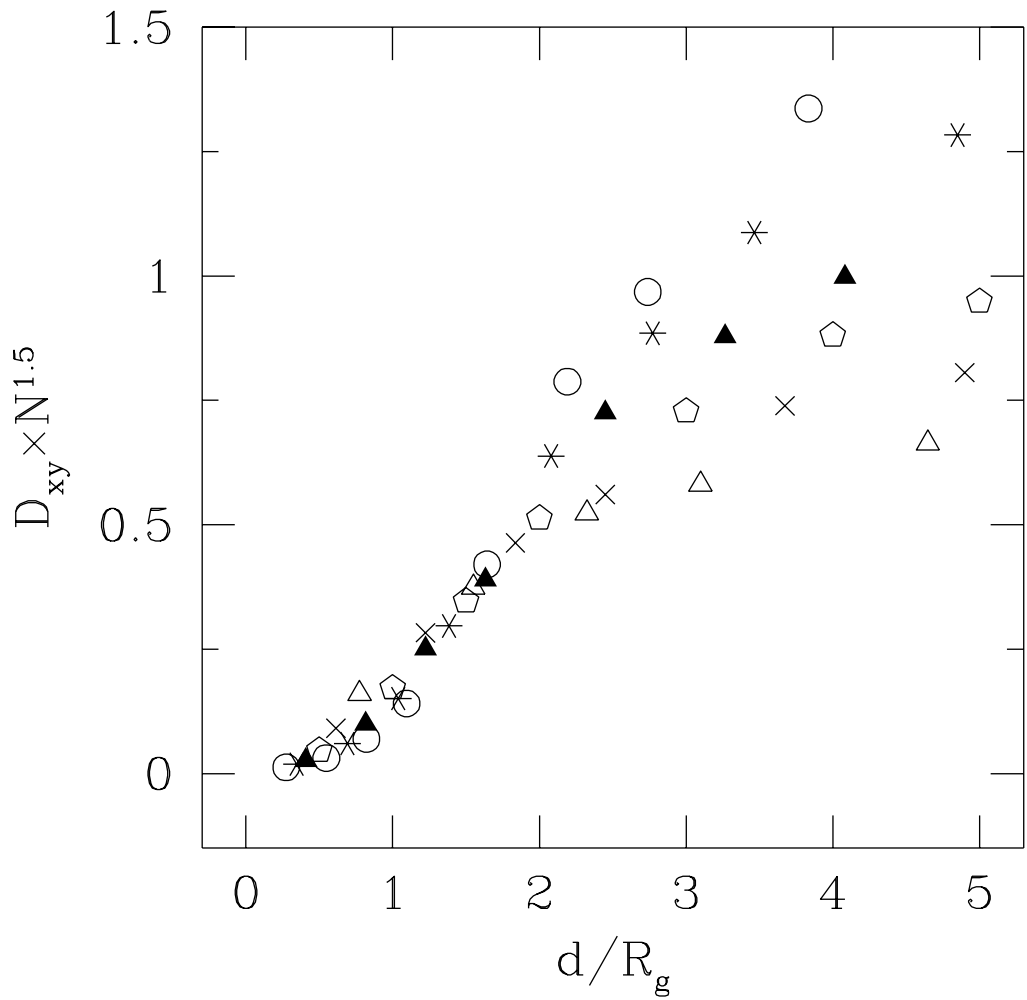,width=6in}
\end{center}
\caption[]{Scaling graph for~$\Dxy$. The symbols represent the same
  $N$'s as in~\fig{scaling_z}. The break-down of the scaling when
  $d>\Rg$ is more dramatic here than in~\fig{scaling_z}.}
\label{fig:scaling_xy}
\end{figure}

The scaling behavior in this regime
is shown in \fig{scaling_z} and \fig{scaling_xy}
for chain lengths up to~80.
As $d$ increase beyond~$\dc$ the scaling clearly breaks down, which is
the topic of the next subsection.

\subsection{$d \gsim \dc$}

In this regime the chain diffuses in
a layered structure of free space and membrane.
Thus depending on the relative importance of each,
the dynamics
of the chain can be either more Rouse-like or more reptation-like.

We can use a simple
resistor-network analogy%
\cite{MDPGdG:SomeRemarks}
to derive the effective
diffusion constants in this regime.
As will be shown shortly $\Dz$ and $\Dxy$ have very different
behavior. 

First we start with~$\Dz$.
We divide the space into two categories; an entanglement zone of
thickness~$\Rg$ and a Rouse zone of thickness~$d-\Rg$.
The effective resistance per length~$d$ in the $z$-direction per unit
($xy$-) area, $\CR$, is the sum of two components,
\begin{equation}
  \CR = \CRE + \CRR
\end{equation}
or
\begin{equation}
  \frac{d}{\Dz} = \frac{\Rg}{\DE} + \frac{d-\Rg}{\DR}
\end{equation}
where $\DR$ is the Rouse diffusion constant and $\DE \sim N^{-3/2}$
is the average diffusion constant in the entanglement zone
as obtained in earlier sections.
Then we obtain
\begin{equation}
  \Dz = \DR\left(1+\frac{\Rg}{d}\frac{(\DR-\DE)}{\DE}\right)^{-1}
\end{equation}
Note that the crossover point from the entangled to 
the Rouse behavior is 
\begin{equation}
  d \approx \Rg\frac{\DR}{\DE} \sim Nb
\end{equation}
contrary to the naive expectation of~$N^{1/2}b$.

The diffusion constant in the $xy$-plane, $\Dxy$, can be obtained in
a similar manner except that the two components divided as above
are connected in parallel rather than in series.

The effective conductance per unit length along a direction
perpendicular to the $z$-axis,
per rectangular area with height~$d$ and unit width, 
denoted by~$\CC$, is again
the sum of two components,
\begin{equation}
  \CC = \CCE + \CCR
\end{equation}
or
\begin{equation}
  d \Dxy = \Rg\DE + (d-\Rg)\DR
\end{equation}
yielding
\begin{equation}
  \Dxy = \DR\left(1-\frac{\Rg}{d}\frac{(\DR-\DE)}{\DR}\right)
\end{equation}
The entangled dynamics cross over to Rouse dynamics 
at~\mbox{$d\sim\Rg$}
differing from the case of~$\Dz$ above.
Hence in the spacing range $N^{1/2}b < d < Nb$, 
we have very anisotropic
diffusion constants.
While the dynamics in the $xy$-plane is almost free the effects of the
membranes are still strong along the $z$-direction.
Beyond this spacing~${}\sim Nb$, the dynamics becomes isotropic
again with Rouse-like diffusion constants in both directions.
The trend is clear from \fig{diffusion}.

The permeability for a stack of $m$~membranes can be defined in a
similar way as in \se{permeability}.
The permeability is the diffusion constant perpendicular
to the slab
divided by its effective thickness
as shown in \eqn{perm2}.
Since the thickness of the stack of $m$ membranes is~$md$
for large~$m$, the
permeability~$P$ is given by
\begin{equation}
  P \sim \frac{\Dz}{md}
\end{equation}
For large~$m$,
$P$ varies as~$N^{-2}$ in the dense regime and $N^{-1}$ in the
dilute regime identical to the scaling of~$\Dz$.

\section{Escape of an entangled chain from a membrane}
\label{sec:escape}

So far we have concentrated on the diffusion of the center of mass of
a chain.
Now we look at the
mean square displacements of a monomer in the middle of
the chain, which
reveals small-scale dynamics in more detail%
\cite{MDSFE86:TheoryPolymer}.

\begin{figure}[p]
\begin{center}
\
\psfig{file=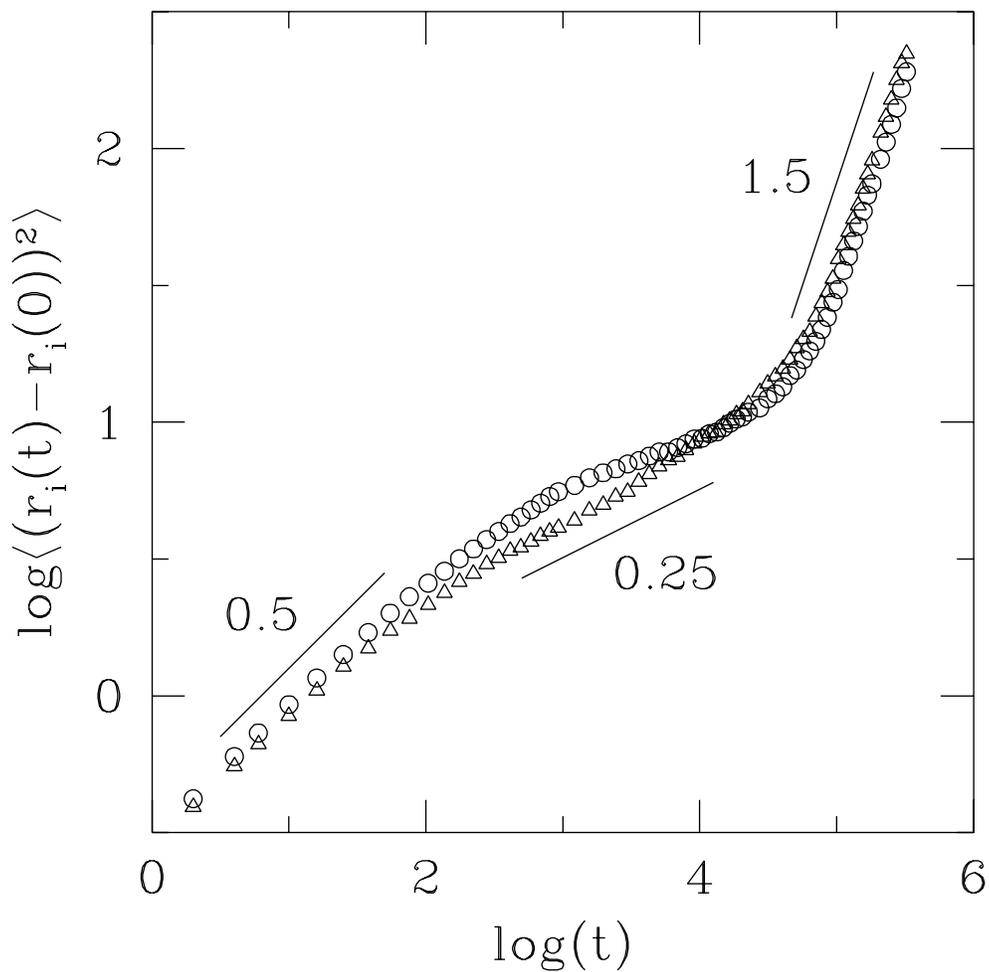,width=6in}
\end{center}
\caption[]{Average mean square displacements of 40 monomers at the
  center of a chain with $N=200$. $r_i$ represents
  $z$-component(triangle) or $x$(or $y$)-component(circle) of  the
  coordinate of a monomer. The curves were averaged over 200 Monte
  Carlo runs. After the $t^{1/4}$ regime $\Dz$ seems to show $t^{1/2}$
  behavior again, but it is not very clear from this graph.} 
\label{fig:escape}
\end{figure}

First we place a chain and a membrane so that 
the chain is maximally entangled, that is,
the center of mass of the chain lies in the membrane.
After the escape time (defined 
in \se{relaxation_time}),
the chain will diffuse from the entanglement
zone into free space.
This process is shown in \fig{escape} for $N=200$.
First we see $t^{1/2}$ behavior up to $N\tau_1$ which 
should be the local
relaxation of Rouse segments, that is, the chain segment 
relaxation between entanglement points.
At greater times
the membrane starts to restrict the motions of chain which makes
the relaxation much slower, probably
proportional to~$t^{1/4}$ as shown in the figure.
In this regime the dynamics in the $z$-direction and $xy$-direction
are different as expected because of the two-dimensional
nature of the membrane.
We need to simulate larger chains to obtain 
the fine structure 
more accurately in this regime.

After the Rouse time scale~$N^2\tau_1$, the chain starts to escape
from the membrane.
This process is unusually fast in the sense that the exponent is
larger than 1, about 3/2 as shown in \fig{escape}.
This is probably because the farther the chain moves from
the membrane the less entangled it is with the membrane,
and hence the larger the diffusion constant.
The chain will eventually escape from the membrane at a time
scale~$N^{5/2}\tau_1$ as obtained in \se{relaxation_time} and
normal diffusion
$\la Z^2 \ra \sim t$ will resume, which is not shown in
the figure.

The monomer relaxation behavior for densely packed
multiple membrane case can be
interpolated from the data shown here and those given by reptation
theory in a three-dimensional network.
See, for example, reference~\cite{MDSFE86:TheoryPolymer}.

\section{Conclusion}
\label{sec:conclusion}

The dynamics of a flexible polymer in the vicinity of 
two-dimensional
networks of topological obstacles shows interesting behavior. 
We have investigated the diffusion constants of a polymer of
polymerization index~$N$ in a solution in which the 
two-dimensional networks are
stacked with uniform spacing~$d$.

Three regimes of dynamics have been shown to exist as a function of
$N$ and~$d$.

\begin{description}
\item[Reptation regime]
When $d$ is much smaller than the radius of gyration of the 
polymer~$N^{1/2}b$, 
the dominant mode of relaxation is reptation and simple
application of de~Gennes's reptation theory reproduces our Monte
Carlo data quite well.
The diffusion constant is proportional to~$N^{-2}$ and to~$d$.

\item[Intermediate regime] 
While the chain is almost free in the direction parallel to the
membrane when $d > N^{1/2}b$,
the motion of the chain perpendicular to the membrane is still
dominated by
membrane penetration up to $d \sim Nb$.
Hence the dynamics of the chain is highly anisotropic in this regime,
the ratio $\Dxy / \Dz$ being $\Rg\DR/(d\DE) \sim N/d$.

\item[Rouse regime] 
When $d$ becomes larger than the chain length~$Nb$ 
the diffusion of the chain is
predominantly Rouse-like and the correction to the Rouse diffusion
constant is 
$\bigO(N^{1/2}\Rg/d)$ for~$\Dz$ and 
$\bigO(\Rg/d)$ for~$\Dxy$.
The penetration time for a single isolated membrane
is found to be proportional to $N^{5/2}$. This can be interpreted as
slowing down of the diffusion of a polymer in the perpendicular
direction to the membranes 
by an extra time of $N^{1/2}\tau_1$ {\em per membrane}.

\end{description}

This study in principle
can provide a new way of filtrating polymers using highly
permeable membranes. The permeability for a single membrane has been
found to be proportional to~$N^{-2}$.
The permeability for multiple membranes will be roughly proportional
to~$\Dz$ divided by the thickness of the stack, providing very rich
behavior depending on the spacings between membranes.

Even though we hope that this study gives impetus to
filtration techniques using fishnet-like membranes, this work
omits important effects.
For example we have completely neglected excluded volume effects,
which could, for instance, make the membranes effectively repulsive to
the polymers.
The effect of membrane fluctuations%
\cite{JAATCL:FluctuationsSolid,RLMG:ShapeFluctuations,%
FFADRN:FluctuationsFlat}
were also neglected in this work.
Even though the time scale of large length scale fluctuations of 
a two-dimensional system is much larger
than that of linear polymer relaxation, it could change some
important characteristics of the filtration process, for instance
by varying the
distances between the membranes, which were assumed uniform in this
paper.

\section*{Acknowledgment}

This work is supported by NSF Grant DMR-9419362 and acknowledgment
is made to the Donors of the Petroleum Research Fund, administered
by the American Chemical Society for partial support of this research.

\end{document}